\documentclass[10pt,a4paper]{article}
\usepackage[dvips]{graphicx}
\usepackage{amssymb}
\usepackage{natbib}
\usepackage{amsmath}
\usepackage{fontenc}
\usepackage[left=3cm,top=3cm,bottom=3cm]{geometry}
\usepackage{textcomp}
\begin{document}

\noindent Electronic version of an article published as \mbox{J. Integr. Neurosci.} 06(02):279--307 (2007), DOI: 10.1142/S0219635207001520 \\
\noindent \textcopyright World Scientific Publishing Company, http://www.worldscientific.com/worldscinet/jin

\newpage

\title{Variability of model-free and model-based quantitative measures of EEG}
\author{S. J. van Albada$^{a,b}$ \and C. J. Rennie$^{a,b}$ \and P. A. Robinson$^{a,b,c}$}
\maketitle

\begin{center}
$^a$School of Physics, The University of Sydney\\
New South Wales 2006, Australia\\
$^b$The Brain Dynamics Centre, Westmead Millennium Institute\\
Westmead Hospital and Western Clinical School of the University of Sydney\\
Westmead, New South Wales 2145, Australia\\
$^c$Faculty of Medicine, The University of Sydney\\
New South Wales 2006, Australia\\


\end{center}

\begin{abstract}
Variable contributions of state and trait to the electroencephalographic (EEG) signal affect the stability over time of EEG measures, quite apart from other experimental uncertainties. The extent of intra-individual and inter-individual variability is an important factor in determining the statistical, and hence possibly clinical, significance of observed differences in the EEG. This study investigates the changes in classical quantitative EEG (qEEG) measures, as well as of parameters obtained by fitting frequency spectra to an existing continuum model of brain electrical activity. These parameters may have extra variability due to model selection and fitting. Besides estimating levels of intra-individual and inter-individual variability, we determine approximate time scales for change in qEEG measures and model parameters. This provides an estimate of the recording length needed to capture a given percentage of the total intra-individual variability. Also, if more precise time scales can be obtained in future, these may aid the characterization of physiological processes underlying various EEG measures. 
Heterogeneity of the subject group was constrained by testing only healthy males in a narrow age range (mean $=$ 22.3 years, sd $=$ 2.7). Resting eyes-closed EEGs of 32 subjects were recorded at weekly intervals over an approximately six-week period. Of these 32 subjects, 13 subjects had follow-up recordings spanning up to a year. QEEG measures, computed from Cz spectra, were powers in five frequency bands, alpha peak frequency, and spectral entropy. Of these, theta, alpha, and beta band powers were most reproducible. Of nine model parameters obtained by fitting model predictions to experiment, the most reproducible ones quantified the total power and the time delay between cortex and thalamus. About 95\% of the maximum change in spectral parameters was reached within minutes of recording time, implying that repeat recordings are not necessary to capture the bulk of the variability in EEG spectra likely to occur in the resting eyes-closed state on the scale of a year.
\end{abstract}

\section{Introduction}

The electroencephalogram (EEG) provides a noninvasive picture of brain electrical activity with a high temporal resolution of the order of milliseconds. As a clinical tool, it has been used successfully in a wide range of contexts, including characterizing sleep disorders, charting the effects of focal abnormalities and epilepsy, and as an aid in the diagnosis of dementia \citep{Niedermeyer1982}. Applications to other disorders, such as attention-deficit hyperactivity disorder (ADHD), post-traumatic stress disorder (PTSD), and depression, have also been investigated.

Measures of EEG are influenced by state and trait contributions, and by factors such as instrumental noise and muscle artifact. Parameters obtained by fitting experimental frequency spectra to predictions from physiology-based models of EEG generation are also affected by the choice of model and fitting procedure, which may or may not provide an optimal and unique solution to the inverse problem. The relative contributions of state, trait, and noise determine the stability of the relevant parameters, which can be quantified by the amount of intra-individual variability. It is essential to know the normal amount of intra-individual variability when determining whether an observed difference between two measurements of the same individual is statistically significant. This in turn may help to estimate the clinical significance of the difference. 

As a vast amount of neuroscientific data is becoming available in standardized databases [see, e.g., \citet{Gordon2005}], candidate markers have emerged for many psychiatric disorders. The specificity of such markers is important in the context of increasingly personalized medicine. Standardization of selection criteria and acquisition methods allows reliable extrapolation of reproducibility and stability findings to individuals within the same database.

Reproducibility, as defined here, depends on the relative amounts of intra-individual and inter-individual variability. Parameters for which the ratio of intra-individual variation to group variation is small distinguish better between individuals than parameters for which this ratio is large. Reproducible parameters are likely to be less affected by noise than others, although they may also have a relatively large variable component due to trait as opposed to state. To estimate the relative contributions of state, trait, and noise, it is therefore necessary to quantify both intra-individual and inter-individual variability. This information is complementary to independent measures, such as correlations with direct physiological measurements, which are not the topic of the current paper. The physiological interpretation of the particular model parameters used here was investigated in detail in \citet{Robinson2004} and related works, including testing for consistency with independent estimates.

Previous work by a number of authors has investigated the variability of the EEG. A sample of these studies reveals a great diversity in focus, study design, and analysis. Results are greatly dependent on whether the variability is compared to the mean, to the variation in the population (as with correlation coefficients), or is taken as a raw quantity. \citet{Oken1988} measured the degree of short-term variability in eyes-closed EEG spectra of adults aged 17--75 years, and found median frequency and peak power frequency to be more stable than absolute or relative band powers (band power divided by total power), where the frequency bands were delta (1.5--4 Hz), theta (4.25--8 Hz), alpha (8.25--13 Hz), beta$_1$ (13.25--20 Hz), beta$_2$ (20.25--32 Hz), and total (1.5--20 Hz). \citet{Hawkes1973} studied the influence of visual tasks on intra-individual EEG variability, which did not appear to have a significant effect. For 27 subjects (16 males, 11 females) in the age range 19--59 years, they found EEG variability to be independent of both age and sex. \citet{Matousek1973} determined coefficients of variation for differences between individuals and within individuals in recordings of children and adolescents aged 1--21 years. Intra-individual variability was found to be highest in the alpha bands (7.5--12.5 Hz) and lowest in the beta$_1$ band (12.5--17.5 Hz). Like Hawkes and Prescott, Matou{\v s}ek and Peters{\'e}n found EEG intra-individual variability to be largely independent of age, while inter-individual variability was smaller in younger subjects. \citet{Gasser1985} and \citet{Salinsky1991} compared within-test to between-test reliability of broad-band spectral parameters using rank correlations. Both studies revealed within-session reliability to be higher than between-session reliability in all bands and at various sites. According to Gasser et al., reproducibility of spectral parameters varied across frequency bands, but not so much across electrodes, while Salinksy et al. found reliability to be generally lower at the T3 and T4 sites. \citet{Fein1983} computed intraclass correlations \citep{Winer1971} for absolute and relative band powers in the eyes-open and the eyes-closed conditions; both showed good reliability, which was comparable for absolute and relative power measures. A more thorough overview of the existing literature is given in the Discussion.

Aging is one possible cause of changes in the EEG. Various studies have shown the high-frequency content of the background EEG to increase with age in healthy subjects \citep{Duffy1984, John1980, Matousek1973} with a concurrent decrease in low-frequency activity \citep{Duffy1984}. Disease or brain damage, on the other hand, tends to shift the alpha peak to lower frequencies \citep{John1977}. Overall EEG amplitude has been observed to be inversely related to age \citep{Matousek1967}. To control the effect of age as a confounding factor, we restrict our study to a limited age range of 18--28 years. The EEG is expected to show little systematic development in this age range, due to the slowing down of changes in brain electrical activity in late adolescence to early adulthood \citep{Matousek1973}. 

On shorter time scales, changes in brain electrical activity may occur due to external stimuli, drowsiness, or changes in attention or processing. All this occurs in an interplay with local or global changes in blood flow and neurotransmitter levels. In addition, various factors contribute to uncertainty in the data, most notably muscle activity, head movement, imperfect calibration, and loss of data due to digitization. We can classify these sources of changes in the EEG into controlled (i.e., fixed or manipulated) variables, such as sex, age, or attention; and uncontrolled (i.e., unmeasured or random) variables, including instrumental noise and head shape. Of course, variables that are controlled in one study may be sources of noise in another study.

\begin{table}[ht]
\centering
\begin{tabular}[ht]{|lll|}
\hline
&\textbf{Sources of Variability}&\\
\hline
 & \emph{Controlled} & \emph{Uncontrolled} \\
\hline
\emph{Static} & Sex & Genotype \\
& Montage &  \\
\emph{Slow (years)} & Age & Chronic anxiety \\
\emph{Slow - Medium (months)} & Menstrual cycle & Seasonal changes, \\
&&Chronic anxiety \\
\emph{Medium (days)} & Diurnal rhythm & \\
\emph{Medium - Fast (hours)} & Caffeine intake & Ultradian rhythms \\
& & Calibration drift \\
\emph{Fast (minutes)} & Attention & Arousal \\
\emph{Very fast (seconds)} & Stimulus-related activation & Instrumental noise\\
& of primary cortex & Muscle artifact \\
 & & Brain microstates \\
\hline 
\end{tabular}
\caption{Controlled (fixed or manipulated) and uncontrolled (random) sources of variability in the EEG on various time scales, and including factors contributing to intra-individual as well as inter-individual variation. A variable that is controlled in one study may be uncontrolled in another; for instance, while seasonal changes may be easily controlled in a short-term study, they were uncontrolled in the present study, which ran over approximately a year. The classification shown here corresponds to the present study.}
\label{sources_of_variability}
\end{table}

An overview of controlled and uncontrolled factors associated with changes in the EEG on various rough time scales is presented in Table \ref{sources_of_variability}. Of these, some contribute to within-subject variability (e.g., attention and brain microstates), others to between-subject variability (e.g., genotype and sex). Most factors contribute to both types of variability. The significance of the time scales of the various uncontrolled sources of variability is that they affect the amount of variability in a recording. For example, a brief recording will have uncertainty due only to instrumental noise, muscle artifact, and multiple microstates, while a recording (or set of recordings) covering a longer span of time will have additional sources of variance. Of course the advantage of longer recordings is usually that the greater volume of data allows for more precise statistical estimation. Balancing these factors is part of experimental design, but this is rarely done quantitatively. 

The aim of the current study is to investigate the variability of spectra both in the `classical' way [qEEG; for an early review see \citet{Duffy1994}] and with reference to a recent physiology-based quantitative model of EEG generation. We reiterate that we focus on stability and reproducibility and that the physiological interpretation and consistency of the model parameters, which were investigated in detail for instance by \citet{Robinson2004, Rowe2004b}, is not at issue in this study.

The comparison between the reproducibility of qEEG measures and that of model parameters provides a valuable testing ground for the utility of the model. There is no reason to assume that model parameters have a larger component of state as opposed to trait than qEEG measures, especially if we compare model parameters and qEEG measures that have a straightforward relation (we will see below that our model has a parameter which is directly related to alpha peak frequency, and one that reflects the total spectral power). Therefore, if the model is a complete and accurate description of the processes responsible for EEG generation, and if the model parameters can be uniquely fitted without being greatly affected by noise in the spectra, we expect reproducibility of model parameters to be comparable to that of qEEG measures. 

Besides estimating overall levels of variability of qEEG measures and model parameters, we consider variability as a function of the time interval between recordings. This allows us to compute approximate characteristic time scales for variability in the various spectral measures, which may be related to underlying processes in future work, using in particular the relationship between the model parameters and brain physiology suggested by our model of EEG generation.

Elaborating on previous work by a number of researchers \citep{Freeman1975, Jirsa1996, Lopes1974, Nunez1974, Nunez1995, Steriade1990, Wilson1973, Wright1996} the model provides a large-scale continuum representation of corticothalamic dynamics by averaging neural properties over a few tenths of a millimeter. Biophysical properties included in the model are rise and decay rates of the potential at the soma, synaptic strengths, nonlinear responses to cell-body potential, axonal ranges and transmission speeds, connectivities between excitatory and inhibitory neural populations in cortex and thalamus, and volume conductivity through the cerebrospinal fluid, skull, and scalp. The model, its steady states, parameter dependences, effects of feedback, and behavior in many regimes have been described and investigated in detail by \citet{Robinson1997, Robinson1998, Robinson2001, Robinson2002, Robinson2003a, Robinson2003b, Robinson2004, Robinson2005}, \citet{Rennie1999}, and \citet{Rowe2004a}. A brief overview is given in Sec.~\ref{sec:theoretical_model}.

\section{Methods}

The study design was longitudinal, following all subjects over a period of about six weeks at approximately one-week inter-test intervals, and some subjects another three or four times over the course of a year. This was done to compare intra-individual variability over time with the inter-individual variation in the population. Consistent time and age trends were expected to be minimal due to the limited age range and duration of the study, and variability to consist mainly of fluctuations around some mean value for each subject.  

\subsection{Subjects}

All 32 subjects were healthy males with an age at the first session in the range of 18 to 27 years (mean $=$ 22.3, sd $=$ 2.7) selected from the Brain Resource International Database (BRID) \citep{Gordon2005}. To control sources of variation in the sample, a number of criteria precluded participation in the study: brain injury, a personal or family history of mental illness, psychological, psychiatric, neurological, or genetic disorders, a personal history of heavy drug or alcohol use, a personal history of cancer or a serious medical condition related to thyroid or heart, a blood-borne illness, or a serious impediment to vision, hearing, or hand movement. In addition, participants in the present study were right-handed, non-smokers, had a mass of 65--95 kg, had no previous exposure to stimulant medication, had not been taking any prescribed medications on a regular basis in the months prior to study entry (except antibiotics, which were completed at least two weeks before the start of the study), and tested negative for illicit drug use at the outset. Participants were also asked to refrain from caffeine consumption for at least three hours before testing, and to forego alcohol for at least 24 hours prior to testing.

\subsection{EEG recording and quantification}

All subjects underwent six consecutive weeks of testing, of which 13 subjects were retested after intervals of three to four months over the course of approximately a year. Allowing for drop-outs, this gave us 208 recordings at intervals of 1--71 weeks. Half of the subjects were tested at Westmead Hospital in Sydney, the other half at Queen Elizabeth Hospital in Adelaide. To limit the effect of diurnal variations, the EEG was recorded at the same hour of day every session, chosen to be between 9 and 10 am for subjects to be awake and non-drowsy. Eyes-closed resting scalp EEG was recorded with a NuAmps amplifier (Neuroscan) at 26 electrode sites according to an extended International 10--20 system at a sampling rate of 500 Hz with average of mastoids as reference. Only the Cz electrode was used in the present analysis, since it is the least affected by muscle artifact. Subjects were seated in a sound and light attenuated room at a constant temperature of $24^{\circ}$C. Skin resistance was less than 5 k$\Omega$. The electro-oculogram was recorded at four channels to correct offline for vertical and horizontal eye movements according to the method of \citet{Gratton1983}. Corrected data were low-pass filtered at 40 dB per decade above 100 Hz. Power spectra for the Cz electrode were computed from two minutes of EEG by multiplying sequential 2.048 s epochs by a Welch window and performing a Fast Fourier Transform with 0.49 Hz resolution.

The following qEEG measures were computed from spectra: (i) band powers: delta (0.2--3.7 Hz), theta (3.7--8.1 Hz), alpha (8.1--12.9 Hz), beta (12.9--30.5 Hz), and gamma (30.5--49.6 Hz). These frequency limits were chosen as the midpoints between frequencies at which the power spectral density was computed (each 0.49 Hz); (ii) total power: the sum of the five band powers; (iii) alpha peak frequency, defined to be that frequency at which the power was maximal; and (iv) spectral entropy \citep{Shannon1948a, Shannon1948b}, defined as
\begin{equation}
H(p_1, p_2, \ldots, p_N) = - \frac{1}{\mathrm{ln} N} \sum_{i=1}^N p_i \mathrm{ln}~p_i,
\end{equation}
where $i$ is a frequency index, $N$ the total number of frequency bins, and $p_i$ is the spectral density, normalized by dividing by the total power. The combination of the natural base for the logarithm and the prefactor $1/\mathrm{ln}\:N$ is chosen so that the entropy obeys $H(1/N,\ldots, 1/N) = 1$, which provides a normalization for the case of equiprobable frequencies (i.e., a flat spectrum).

The spectral entropy is a measure of the structure of the spectrum. It is greater when power is distributed evenly over a large range of frequencies than when power is concentrated mainly in a few peaks. This corresponds with the intuitive meaning of entropy as a measure of the number of active degrees of freedom in a system, since a lower spectral entropy implies a more limited range of frequencies at which the energy of the system resides. The power is thus interpreted as a probability that the energy of the system is concentrated at a certain frequency. Normalization ensures that the entropy depends only on the shape of the spectrum, and not on the total power. It should be noted that the frequencies at which the peaks occur do not matter. Hence, spectral entropy does not say anything about the relative amounts of fast and slow activity. The entropy changes with the degree of activation of the brain, being generally lower with increased drowsiness, but the relationship is not completely straightforward \citep{Jantti2004}. \citet{Inouye1991} noted an increase in spectral entropy mainly in the left hemisphere relative to the resting condition when subjects performed mental arithmetic.

\subsection{Theoretical model}\label{sec:theoretical_model}

\begin{figure}[ht]
\centering
  \includegraphics[width=200pt]{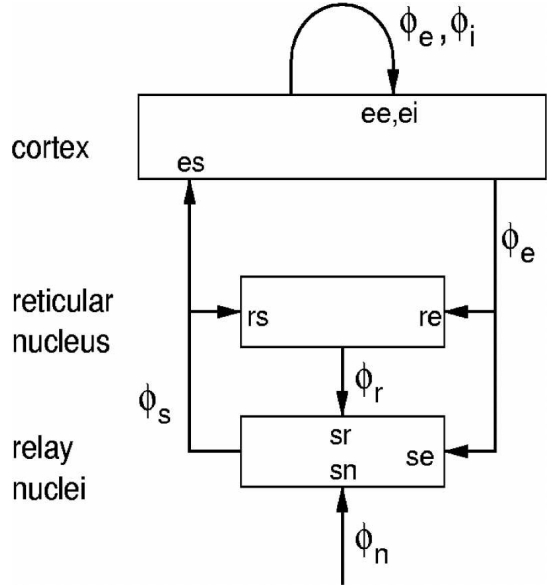}
\caption{Schematic representation of the model components and their interconnections: $e=$ cortical excitatory, $i=$ cortical inhibitory, $s=$ specific relay, $r=$ thalamic reticular. Input from underlying structures is represented by $\phi_n$.}
\label{fig:model_dia}
\end{figure}

The components described by our physiology-based mean-field model of EEG generation are excitatory and inhibitory neurons in the cerebral cortex, excitatory specific relay neurons in the thalamus, and the neurons of the thalamic reticular nucleus, which have an inhibitory effect on their targets. In addition, the model allows for input $\phi_n$ to the thalamus from underlying neural structures, which in the present study was set to have a constant modulus and a random phase in the Fourier domain, corresponding to spatiotemporal white noise. The connections between components are as shown in Fig.~\ref{fig:model_dia}, where $\phi_a$ ($a=e,i,n,r,s$) symbolizes the relevant firing rate field. Long-range corticothalamic connections are excitatory in nature, while inhibitory feedback occurs internally within the cortex and thalamus. Gains are represented in Fig.~\ref{fig:model_dia} by pairs of letters, the first of which refers to the receiving neural population, while the second refers to the type of incoming neurons.

The first equation of the model relates the mean firing rate $Q_a(\textbf{r},t)$ of each population of neurons to the cell-body potential $V_a(\textbf{r},t)$ relative to resting:
\begin{equation}
Q_a(\textbf{r},t) = S[V_a(\textbf{r},t)],
\end{equation}
with $a=e,i,r,s$. Here, $\textbf{r}$ parameterizes the cortex, which is approximated as two-dimensional owing to its relative thinness. The function $S[V_a(\textbf{r},t)]$ has a sigmoidal shape that results from averaging over a large number of step functions representing different threshold responses. It increases smoothly from $0$ to the maximum attainable firing rate $Q_{\rm{max}}$ ($\sim$250 Hz) as $V_a$ runs from $-\infty$ to $\infty$, and takes the specific form
\begin{equation}\label{fun:sigmoid}
S(V_a) = \frac{Q_{\rm{max}}}{1+\mathrm{exp}[-(V_a-\theta)/\sigma']},
\end{equation}
where $\theta$ is the mean threshold potential, and $\sigma' \pi/\sqrt{3}$ is the standard deviation of the distribution of firing thresholds in the neural population. 

The cell-body potentials $V_a$ are made up of contributions from all afferent neurons, whose inputs drive synaptic changes that result in signals that propagate down the dendritic tree. The effect of a given incoming firing rate field $\phi_b$ on the cell-body potential of target neurons of type $a$ depends on the number of synapses from afferent axons onto dendrites of the target population ($N_{ab}$), as well as the typical change in cell-body potential per unit input ($s_{ab}$). The dendritic tree and synapses act in effect as a low-pass filter that attenuates high-frequency activity due to differential delays for signals traveling through them. We use an analytic form for the synaptodendritic filter function which has been found to conform closely to experimental observations \citep{Freeman1991, Rennie1999}. This leads to \citep{Robinson1997, Robinson2004}
\begin{equation}
D_a(t)V_a(\textbf{r},t) = \sum_b N_{ab}s_{ab}\phi_b(\textbf{r}, t-\tau_{ab}),
\end{equation}
\begin{equation}
D_a(t) = \frac{1}{\alpha\beta}\frac{d^2}{dt^2} + \left(\frac{1}{\alpha}+\frac{1}{\beta}\right)\frac{d}{dt} + 1.
\end{equation}
Here, $\tau_{ab}$ represents any discrete time delay for signals to travel from neurons of type $b$ to neurons of type $a$. As an approximation, the discrete time delays for signals traveling between neural populations are taken to be zero locally within the cortex and thalamus. The only nonzero $\tau_{ab}$ are therefore $\tau_{es}=\tau_{se}=\tau_{re}=t_0/2$, with $t_0$ the time necessary to complete a full loop through cortex and thalamus. The inverse rise and decay times for the potential at the soma due to dendritic propagation and temporal smoothing are represented by $\beta$ and $\alpha$, respectively. In practice, we set $\alpha=\beta/4$ in rough agreement with experiment \citep{Rowe2004a}, to reduce the number of independent parameters and hence the robustness of those that are independently fitted. The product $N_{ab}s_{ab}$ times the derivative of the sigmoid (\ref{fun:sigmoid}) at the steady-state value of the potential, $V_a^{(0)}$, results in a set of gains, 
\begin{eqnarray}\label{eq:gains_and_rho}
G_{ab} &=& \rho_a N_{ab}s_{ab},\\
\rho_a &=& \frac{dQ_a(\mathbf{r},t)}{dV_a(\mathbf{r},t)}\Big{|}_{V_a^{(0)}}, 
\end{eqnarray}
which characterize the connection strengths between neural populations. 
 
The final aspect of the model is axonal propagation that spreads electrical activity within the cortex, and can be approximated by a damped-wave equation with source $Q_a(\textbf{r},t)$ \citep{Robinson1997, Robinson2004},
\begin{equation}
\frac{1}{\gamma_a^2}\left[\frac{\partial^2}{\partial t^2}+2\gamma_a\frac{\partial}{\partial t}+\gamma_a^2-v_a^2\nabla^2 \right]\phi_a(\textbf{r},t) = Q_a(\textbf{r},t).
\end{equation}
Here, $\nabla^2$ is the Laplace operator, representing a second-order spatial derivative. The damping rate $\gamma_a=v_a/r_a$ is the ratio of the average axonal transmission speed $v_a \approx$ 10 m/s and characteristic axonal range $r_a$. In the local inhibition approximation \citep{Robinson1998}, all inhibitory axons are taken to be short enough to justify setting $\gamma_i = \gamma_r = \infty$. Furthermore, local excitatory connections within the thalamus are assumed short enough to set $\gamma_s = \infty$, so that the corresponding equations simply become $\phi_{i,r,s}(\textbf{r},t) = Q_{i,r,s}(\textbf{r},t)$. From now on, we will refer to $\gamma_e$ as $\gamma$, the only spatial damping rate retained. We assume the EEG signal to be proportional to the cortical excitatory firing rate field $\phi_e$, since macroscopic cortical potentials are thought to arise from the activity of many long-range excitatory cortical neurons firing in synchrony. The calculation of the spectral power from $\phi_e$ is quite mathematically involved; for details see the Appendix. The effect of volume conduction on the resulting scalp potential is incorporated in Eq. (\ref{eq:pow}) via a spatial smoothing function. For our present purposes, it suffices to give a brief description of the model parameters that are determined by fitting theoretically predicted spectra to experimental spectra.

In all, we fit eight physiological parameters representing aspects of brain function that are central to the generation of the EEG, in addition to an attenuating factor $p_0$, which represents filtering of the signal through cerebrospinal fluid, skull, and scalp [see Eq. (\ref{eq:P0eq})]. The time delay for signals to complete one full loop between cortex and thalamus via long-range excitatory connections is given by $t_0$. The quantities $\gamma$ and $\alpha$ are inverse time constants; $\gamma$ representing the average damping rate due to spreading cortical activity, and $\alpha$ the decay rate of the potential at the soma. The remaining five parameters are dimensionless gains; $G_{ee}$ representing the average connection strength between excitatory neurons in the cerebral cortex, $G_{ei}$ the average connection strength between short-range inhibitory cortical neurons and their excitatory target neurons, $G_{ese}=G_{es}G_{se}$ the gain for the direct loop from the cortex to the specific relay nuclei of the thalamus and back to the cortex, $G_{esre}=G_{es}G_{sr}G_{re}$ the gain for the same loop but passing through the reticular thalamic nucleus, and finally $G_{srs}=G_{sr}G_{rs}$, the gain for the intrathalamic loop from specific relay nuclei through the reticular thalamic nucleus, and back to the relay nuclei. These are the only independent gains in our model. The gains depend on $\theta$ and $\sigma'$ [cf. Eqs (\ref{fun:sigmoid}) and (\ref{eq:gains_and_rho})], but these are not determined separately.
Approximate nominal values of the independently fitted model parameters are given in Table \ref{table:parameters}. The given parameter ranges are physiologically realistic, in agreement with independent estimates \citep{Robinson2004}.

\begin{table}[ht]
\centering
\begin{tabular}[ht]{|lll|}
\hline
\textbf{Parameter} & \textbf{Description} & \textbf{Range} \\
\hline
$t_0$ & Delay time for the corticothalamic loop & 0.06 -- 0.13 s\\
$p_0$ & Logarithm of an overall multiplicative factor &\\
& for the power & No limits$^*$\\
$\gamma$ & Damping rate of action potentials in the axon & 40 -- 280 $\rm{s}^{-1}$\\
$\alpha$ & Inverse decay time of the cell-body potential & 10 -- 200 $\rm{s}^{-1}$\\
$G_{ee}$ & Gain for interconnections between excitatory &\\
&cortical neurons & 0 -- 20\\
$G_{ei}$ & Gain for inhibitory cortical neurons synapsing & \\
& on excitatory cortical neurons & $-$35 -- 1\\
$G_{ese}$ & Product $G_{es}G_{se}$, i.e., the gain of the direct & \\
& corticothalamic loop & 0 -- 20\\
$G_{esre}$ & Product $G_{es}G_{sr}G_{re}$, i.e., the gain of the loop & \\
& from cortex through the reticular thalamic  & \\
& nucleus, to the specific relay nuclei, and back &\\
& to cortex & $-$30 -- 2\\
$G_{srs}$ & Product $G_{sr}G_{rs}$, i.e., the gain of the  &\\
&intrathalamic loop & $-$15 -- 0.5\\
\hline 
\end{tabular}
\caption{Description of the model parameters with the physiologically realistic ranges \citep{Robinson2004} over which they were allowed to vary. $^*$~Although no limits were imposed on $p_0$, it is possible to calculate an approximate range in which $p_0$ is expected to fall, as shown in the Appendix.}
\label{table:parameters}
\end{table}

\subsection{Model fits}

Parameter estimates were obtained by fitting the logarithm of the theoretical power spectral density, log$(P_{the})$ to the logarithm of the experimental spectral density, log$(P_{exp})$. Taking the logarithm ensures that differences in power spectral density are weighted more or less equally at each frequency. Best fits were obtained by minimizing the $\chi^2$ error between log$(P_{the})$ and log$(P_{exp})$, defined as a sum over all frequencies indexed by $i$, and weighted using the standard deviation $\sigma_i$,
\begin{equation}
\chi^2 = \sum^N_{i=1} \frac{[\mathrm{log}(P_{exp,i}) - \mathrm{log}(P_{the,i})]^2}{\sigma_i^2},
\end{equation}
using a Levenberg-Marquardt optimization \citep{Press1992, Rowe2004a}. Limits corresponding to the ranges in Table \ref{table:parameters} were employed to keep parameters within physiologically realistic ranges. The stopping criterion was that $\Delta \chi^2$ be smaller than $10^{-5}$ for six successive iterations.

The fitting procedure led to the convergence of model parameters in 194 cases out of 208 (i.e., 93.3\%). In the remaining 14 cases, no combination of parameters gave a prediction that was sufficiently close to the experimental spectrum for six successive iterations, even though these spectra were reasonably free of artifact. A single spectrum with $\chi^2 =$ 584.1 was removed because visual inspection revealed a high level of artifact. This left 193 spectra for further analysis, of which $\chi^2$ ranged from 6.6 to 199.5 (mean $=$ 49.3).

Some examples of fits to experimental spectra with varying goodness of fit for both the eyes-open and the eyes-closed state can be found in \citet{Rowe2004a}.

\subsection{Statistical analysis}\label{sec:stats}

The analysis consisted of four parts: 

(1) Means and standard deviations of qEEG measures and model parameters over all trials were computed, and distributions were plotted as histograms. Variability was split into components arising from intra-individual and inter-individual differences, by giving standard deviations due to both sources. Within-subject standard deviations were computed by adding the squared deviations from the mean for each person, then dividing by the number of degrees of freedom, which is the number of trials minus the number of subjects (i.e., $193-32=161$), and finally taking the square root. Between-subject standard deviations were computed to be the root mean square deviation of subject means from the mean of all observations. The total sum of squared deviations is composed of the within-subject and between-subject sums of squares.

Since skewness can reduce the test-retest reliability for otherwise normal data \citep{Dunlap1994}, all quantities were transformed towards the Gaussian with a method based on a transformation described by \citet{Rosenblatt1952} for the following parts of the analysis \citep{vanAlbada2007}. This method yields the best possible agreement with the normal distribution starting from non-normally distributed quantities, and can thus also be applied to quantities for which no standard transformation is known to achieve approximate normality.

(2) Elaborating on intra-individual and inter-individual differences given in Part (1), we calculated ratios of intra-individual to total sample variances, which are related to the distinguishing power of parameters in the normal population. The width of a subject's distribution relative to the population distribution determines the certainty with which we can conclude whether a subject is an outlier on a given measure.

(3) We computed average Spearman rank correlation coefficients [see, e.g., \citet{Becker1988}] between spectral measures from the first six weeks at one-week intervals. Trial-to-trial correlations were determined from 31 recordings in Week 1, 25 in Week 2, 26 in Week 3, 23 in Week 4, 21 in Week 5, and 24 in Week 6. Correlation coefficients provide a direct estimate of the certainty with which future parameter values can be predicted from measured values, and also characterize the relative amounts of intra- and inter-individual differences. Spearman rather than Pearson correlations were used because the rank correlation is the same before and after transformation to the normal distribution.

\begin{figure}[ht]
\centering
\includegraphics[width=280pt]{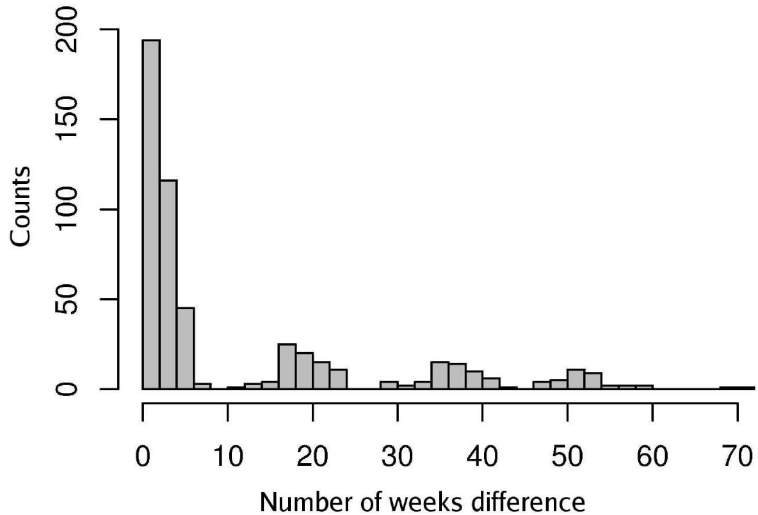}
\caption{Histogram of time intervals between sessions, of length a week or more. For instance, if a subject's EEG was obtained in Weeks 1, 2, 3, and 6, we would count two intervals of a week, one of two weeks, and one each of three, four, and five weeks.}
\label{fig:intervals}
\end{figure}

(4) Finally, we explored the time scales on which the bulk of the intra-individual variation occurred, using the two widely different scales at our disposal: seconds or minutes within trials, and weeks or months between trials. This was done by computing average absolute intra-individual differences in normalized qEEG measures and model parameters for spectra that were a certain number of weeks apart (Fig.~\ref{fig:intervals}), and between spectra from the same trial separated by 30, 60, or 90 s. By normalizing the parameter distributions, intra-individual variation was expressed as a fraction of the population variance. An implicit assumption was that the amount of difference between recordings depended only on the time interval, not on the date of recording itself.

With each time scale is thus associated an average level of change in spectral parameters, which is expected to increase with time up to a certain stage, and then level off when the bulk of intra-individual variation has been sampled. Such a relation between intra-individual differences and the time interval was also noted by \citet{John1987} for intervals of one hour to 2.5 years. Different quantities were expected to change on different time scales, depending on the neural processes and noise contributions that determine them. 

 To equate levels of power and noise in spectra used to assess long-term variability and spectra used to assess short-term variability, all spectra in this analysis phase were computed from 30 s of EEG. For intervals of a week or more, spectra were derived from the third 30 s epoch from each trial---the epoch that was successfully fitted to the model in the highest percentage of cases. The use of such short epochs may result in slightly larger differences between spectra than would be obtained on the basis of longer recordings, although 40 or 60 s epochs were found by some authors to be not significantly more reproducible than 20 s epochs \citep{Gasser1985, Mocks1984}.

Of 193 two-minute recordings, spectra for the first 30 s were successfully fitted to the model in 182 cases, in 178 cases for the next 30 s, in 188 cases for the penultimate 30 s, and in 181 cases for the final 30 s. The number of differences that could be derived was constrained by the number of subsequent spectra coming from the same trial, being $n=485$ in number for spectra taken 30 s apart, $n=328$ for spectra taken 60 s apart, and $n=163$ for spectra 90 s apart.

All measures were transformed to the standard normal distribution for direct comparison of levels of variability. We then calculated median absolute differences in normalized quantities over the relevant time intervals. The median rather than the mean was used as an appropriate measure of central tendency for the highly skewed distributions of absolute differences.

To characterize the magnitude and to derive time scales for changes in the various quantities, we fitted average differences to curves of the form 
\begin{equation}\label{curve}
\Delta(t) = c\left(1-e^{-t/\tau} \right),
\end{equation}
where $\Delta(t)$ is the median absolute difference between spectral parameters after a time interval $t$, $c$ is the asymptotic value of the median absolute difference, and $\tau$ is a time scale for change in the relevant parameter. In general, $c$ will be well determined from the asymptotic level of trial-to-trial differences, but $\tau$ will not be well determined unless it falls near one or the other end of the range of time scales. Still, the fits (\ref{curve}) constrain the possible values of $\tau$, which is useful for estimating the length of EEG trials that are likely to capture a certain percentage of the total intra-individual variability, for the design of longitudinal studies, and for relating EEG variability to underlying factors. Significance values for the goodness of fit to empirical differences were determined by computing the linear correlation between median absolute differences and the function (\ref{curve}).

The above measures of variability complement each other, together allowing in principle for: (i) estimation of the width of individuals' distributions for qEEG measures and model parameters (ii) estimation of the certainty with which we can make predictions about future spectra (correlation coefficients); (iii) estimation of the discriminatory power of spectral measures in the normal population (ratios of intra-individual to population variances); (iv) estimation of the time scale on which changes in brain electrical activity occur, where different scales point to different underlying neural and noise processes (absolute differences as a function of time interval); (v) insight into the physiological sources of intra- and inter-individual variability in EEG (model parameters); and (vi) comparison between intra- and inter-individual variability in qEEG measures and physiologically based parameters.

\section{Results}

Presented here are the findings from the four stages of our statistical analysis: distributions of qEEG measures and model parameters, ratios of intra-individual variances to the total sample variances, trial-to-trial correlations, and intra-individual differences versus the time interval. None of the investigated measures showed a significant time trend over the duration of the study when correcting for multiple comparisons.

\subsection{Distributions}

\begin{figure}[ht]
\centering
  \includegraphics[width=400pt]{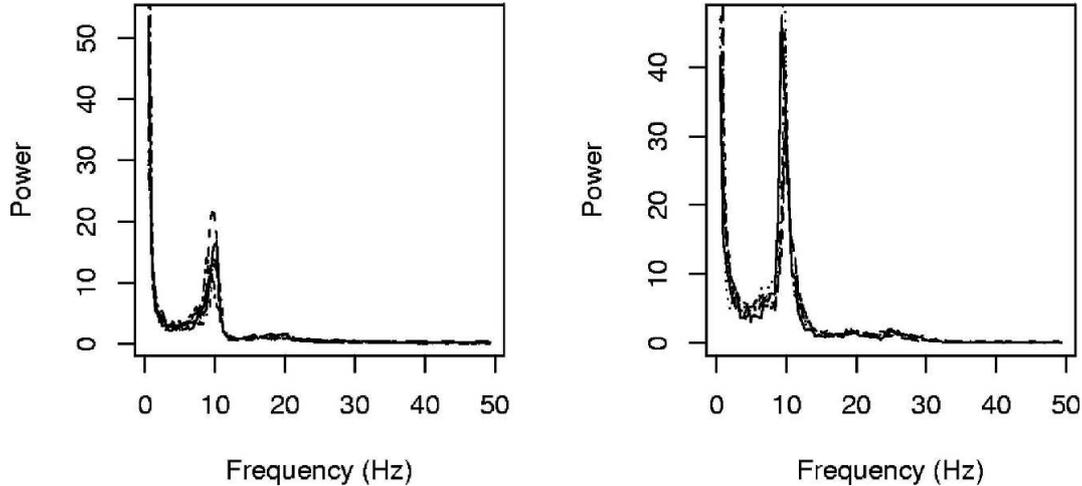}
\caption{Spectra across recording sessions for two subjects at weekly intervals, with power in $\mu \rm{V^2/Hz}$. There are six recordings for the subject on the left (Weeks 1 through 6), and five recordings for the subject on the right (Weeks 1, 2, 3, 4, and 6).}
\label{fig:intra_ex}
\end{figure}

To give an idea of the level of variability in spectra across recordings, spectra for two subjects are plotted in Fig.~\ref{fig:intra_ex}. The two sets of spectra appear quite distinctive, especially in the alpha frequency range.

\begin{table}[ht]
\centering
\begin{tabular}[ht]{|cllllll|}
\hline
&\textbf{Quantity} & \textbf{Unit} & \textbf{Mean} & \textbf{SD} & \textbf{SD} & \textbf{SD} \\
&&&&&\textbf{Within}&\textbf{Between}\\
\hline 
& delta power & $\mathrm{\mu V^2}$ &  190  & 150   & 120   & 250 \\
& theta power & $\mathrm{\mu V^2}$ & 80   & 60  & 20  & 150  \\
& alpha power & $\mathrm{\mu V^2}$ & 110 & 70  & 30 & 150  \\
\textsl{qEEG} & beta power &  $\mathrm{\mu V^2}$ & 34  & 18  & 9  & 41\\
& gamma power &  $\mathrm{\mu V^2}$ & 8  & 7 & 5  & 12 \\
& total power &   $\mathrm{\mu V^2}$ & 420 & 220 & 140  & 440 \\
& alpha peak freq & Hz & 9.3  & 0.8  & 0.5  & 0.7  \\
& entropy & -- &  0.69  & 0.07  & 0.06 & 0.12 \\
\hline 
& $t_0$ & s & 0.079 & 0.010 & 0.008 & 0.017 \\
& $p_0$ & -- & 3.0 & 0.5 & 0.3 & 1.0 \\
& $\gamma$ & $\mathrm{s^{-1}}$  &  80  & 30 & 30 & 40  \\
& $\alpha$ & $\mathrm{s^{-1}}$  & 70  & 30  & 30  & 60  \\
\textsl{Model} & $G_{ee}$ & -- & 4 & 4 & 3 & 6 \\
& $G_{ei}$ & -- & $-$8 & 3 & 3 & 4 \\
& $G_{ese}$ & -- & 12 & 5 & 4 & 9 \\
& $G_{esre}$ & -- & $-$6 & 4 & 3 & 8 \\
& $G_{srs}$ & -- & $-$0.4 & 0.5 & 0.4 & 1.0 \\
\hline
\end{tabular}
\caption{Means and standard deviations of measures derived from experimental and fitted spectra. The last two columns show the partitioning of the sum of squares into within-subject and between-subject contributions. The numbers of significant figures are chosen such that the standard deviation is reasonably small compared to the last significant digit in the mean (no larger than 25 times 10, 1, 0.1, 0.01, or 0.001, depending on the precision of the mean).}
\label{means_sds}
\end{table}

Quantitative analysis of variability in qEEG measures and model parameters shows more clearly the distinctiveness of individual spectra. Table \ref{means_sds} lists means and standard deviations when all data are pooled, along with purely within-subject and between-subject standard deviations.

\begin{figure}[htb]
\centering
\includegraphics[width=400pt,height=350pt]{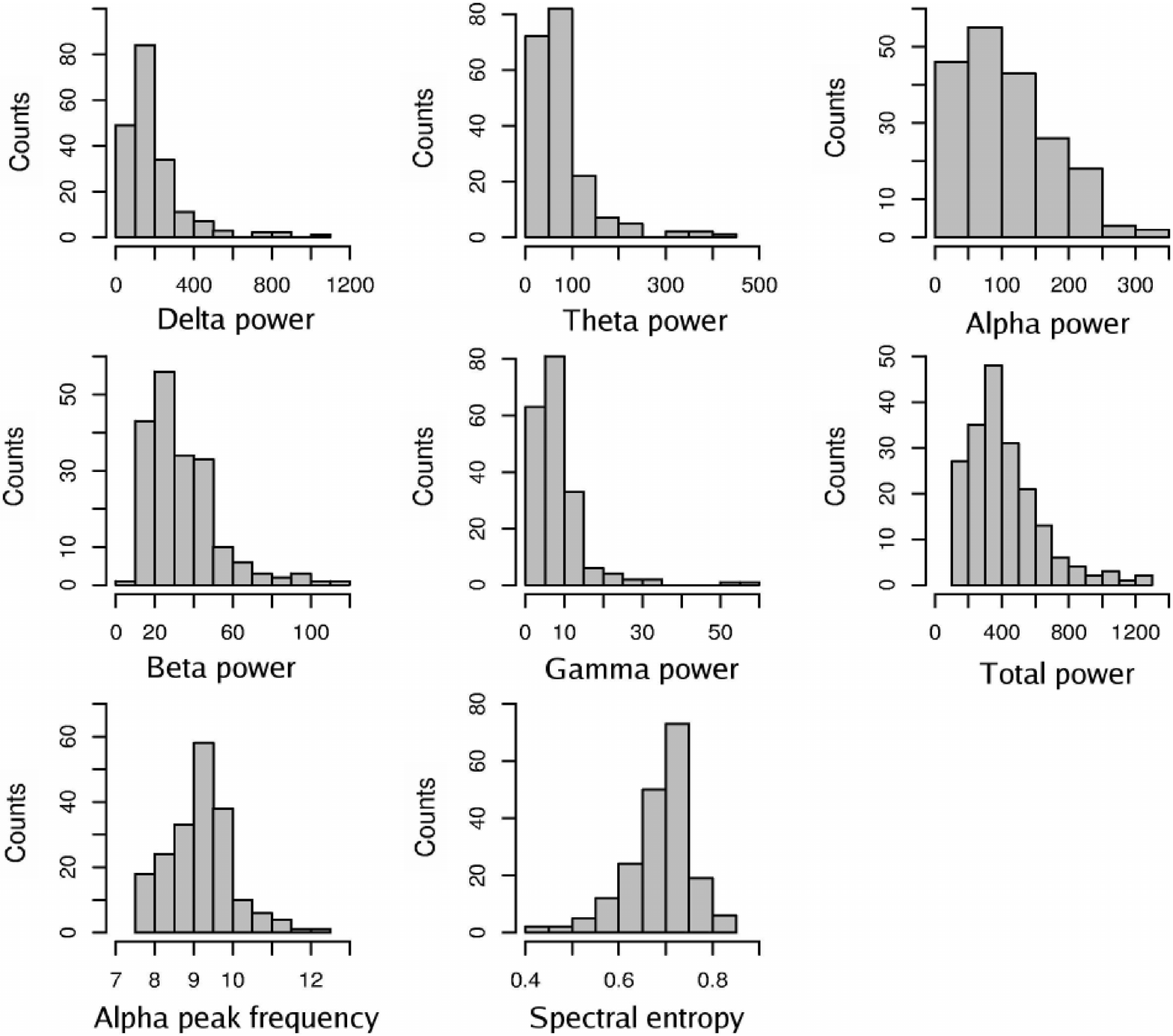}
\caption{Histograms for qEEG measures, giving the number of occurrences versus the value of the measure for 193 spectra. Units for band powers are $\mu \mathrm{V^2}$, the alpha peak frequency is given in Hz, and spectral entropy is dimensionless.}
\label{fig:trad_dist}
\end{figure}

\begin{figure}[htb]
\centering
\includegraphics[width=400pt,height=350pt]{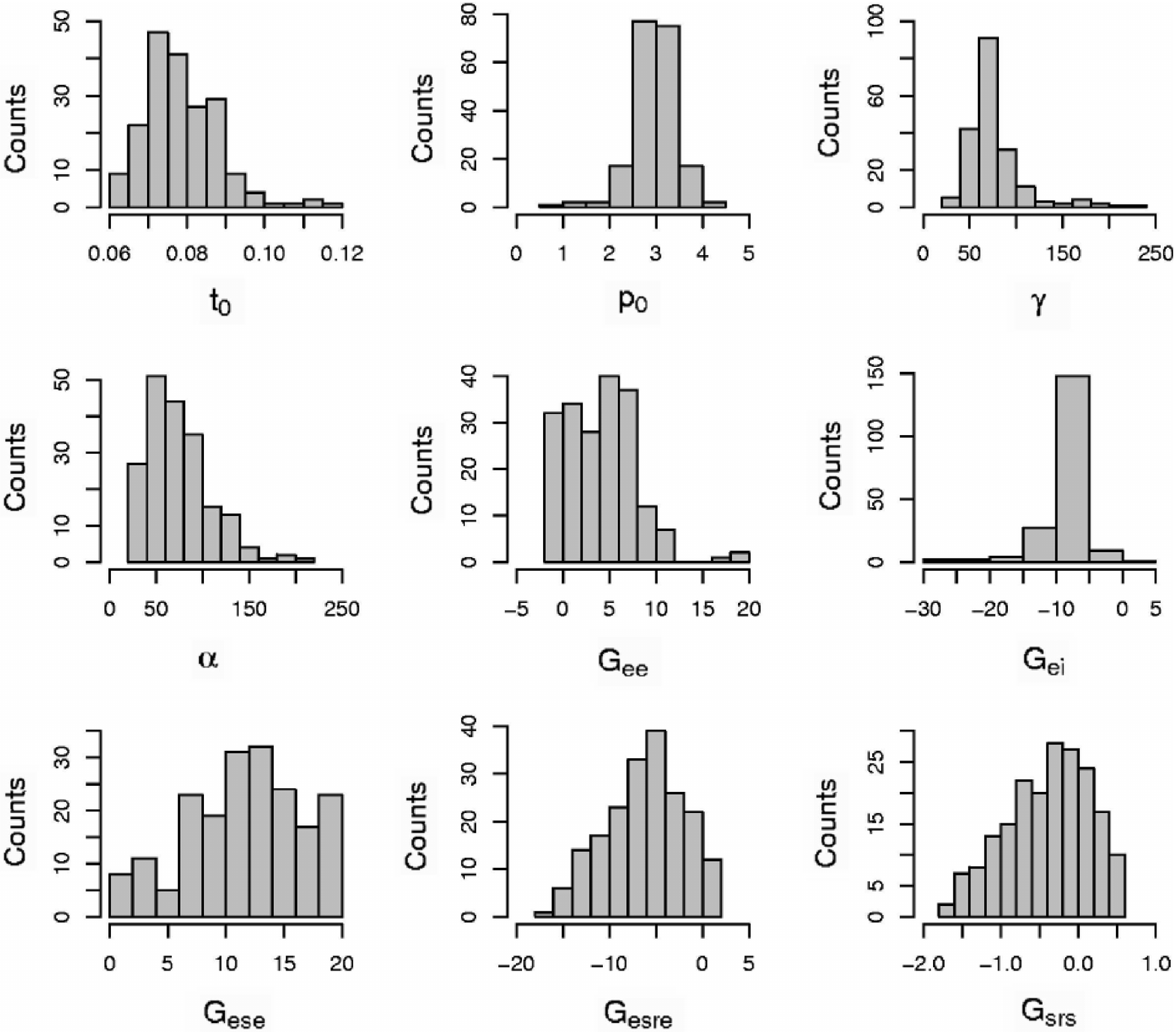}
\caption{Histograms for model parameters, giving the number of occurrences versus the parameter value for 193 spectra. Corticothalamic delay parameter $t_0$ is measured in s, axonal and dendritic rates $\gamma$ and $\alpha$ in s$^{-1}$, and the other parameters are dimensionless.}
\label{fig:param_distr}
\end{figure}

Frequency histograms for qEEG measures are plotted in Fig.~\ref{fig:trad_dist}, while Fig.~\ref{fig:param_distr} shows the distributions of fitted model parameters for the 193 spectra. It is seen that some parameters appear to reach a lower or upper boundary. For instance, the cortical excitatory gain $G_{ee}$ often has a value close to 0, while the gain for the direct corticothalamic loop, $G_{ese}$, is often close to 20. It was investigated whether this was due to constraints imposed during the fitting process, by fitting the model to eyes-closed Cz spectra of 100 healthy subjects recorded at Brain Resource, Sydney (http://www.brainresource.com) while allowing all gains to vary without bound. This resulted in a distribution of $G_{ee}$ very similar to that observed in the present study with bounds, having many values close to 0 but still none smaller than $-1$. We thus interpret $G_{ee} \gtrsim 0$ as a physiological constraint rather than an artificial one imposed during the fitting routine. Similarly, $G_{ese}$ only exceeded 20 in 10\% of cases, and was larger than 30 in only 4\% of cases.

\subsection{Ratios of intra-individual to sample variance}\label{sec:ratios}

Proportions of the total variance accounted for by differences within subjects over time, for both untransformed quantities and quantities transformed toward the standard normal distribution (cf. Sec.~\ref{sec:stats}), are summarized in Table \ref{variance_ratios}. Of transformed qEEG measures, theta, alpha, and beta band power had the lowest ratios of intra-individual variance to the total sample variance (17\%, 19\%, and 25\%, respectively). The delta and gamma bands were less reliable with variance ratios of 49\% and 53\%, respectively, consistent with \citet{Gasser1985}. The spectral entropy was the least reproducible of the investigated qEEG measures with variance ratio 62\%.

\begin{table}[ht]
\centering
\begin{tabular}[ht]{|clccc|}
\hline
&\textbf{Quantity} & \textbf{VR (original)} & \textbf{VR (transformed)} & $\rho$ \\
\hline 
&delta power & 0.64 & 0.49 & 0.45 \\
&theta power & 0.14 & 0.17 & 0.83 \\
&alpha power & 0.20 & 0.19 & 0.87 \\
\textsl{qEEG} & beta power & 0.24 & 0.25 & 0.77 \\
&gamma power & 0.58 & 0.53 & 0.50 \\
&total power & 0.40 & 0.33 & 0.70 \\
&alpha peak freq. & 0.40 & 0.37 & 0.67 \\
&entropy & 0.68 & 0.62 & 0.36 \\
\hline 
&$t_0$ & 0.61 & 0.52 & 0.56 \\
&$p_0$ & 0.40 & 0.44 & 0.53 \\
&$\gamma$ & 0.80 & 0.68 & 0.42 \\
&$\alpha$ & 0.58 & 0.46 & 0.48 \\
\textsl{Model} & $G_{ee}$ & 0.58 & 0.52 & 0.36 \\
&$G_{ei}$ & 0.87 & 0.77 & 0.21 \\
&$G_{ese}$ & 0.58 & 0.61 & 0.41 \\
&$G_{esre}$ & 0.55 & 0.57 & 0.50 \\
&$G_{srs}$ & 0.44 & 0.47 & 0.40 \\
\hline
\end{tabular}
\caption{Ratios of intra-individual to sample variances for qEEG measures and model parameters, next to average Spearman rank correlations ($\rho$) between quantities from Week 1 and quantities from the following five weeks (cf. Sec.~\ref{sec:correlations}). Original quantities are given as well as those transformed to conform to the standard normal distribution (cf. Sec.~\ref{sec:stats}). Rank correlations are the same for both, since all transformations were monotonically increasing. A low ratio of variances corresponds to a high correlation coefficient.}
\label{variance_ratios}
\end{table}

The most distinguishing model parameter was $p_0$, which quantifies the total power, with a ratio of 44\%. Least reproducible was the cortical inhibitory gain $G_{ei}$, with a variance ratio of 77\%.

\begin{figure}[htp]
\centering
\includegraphics[width=400pt]{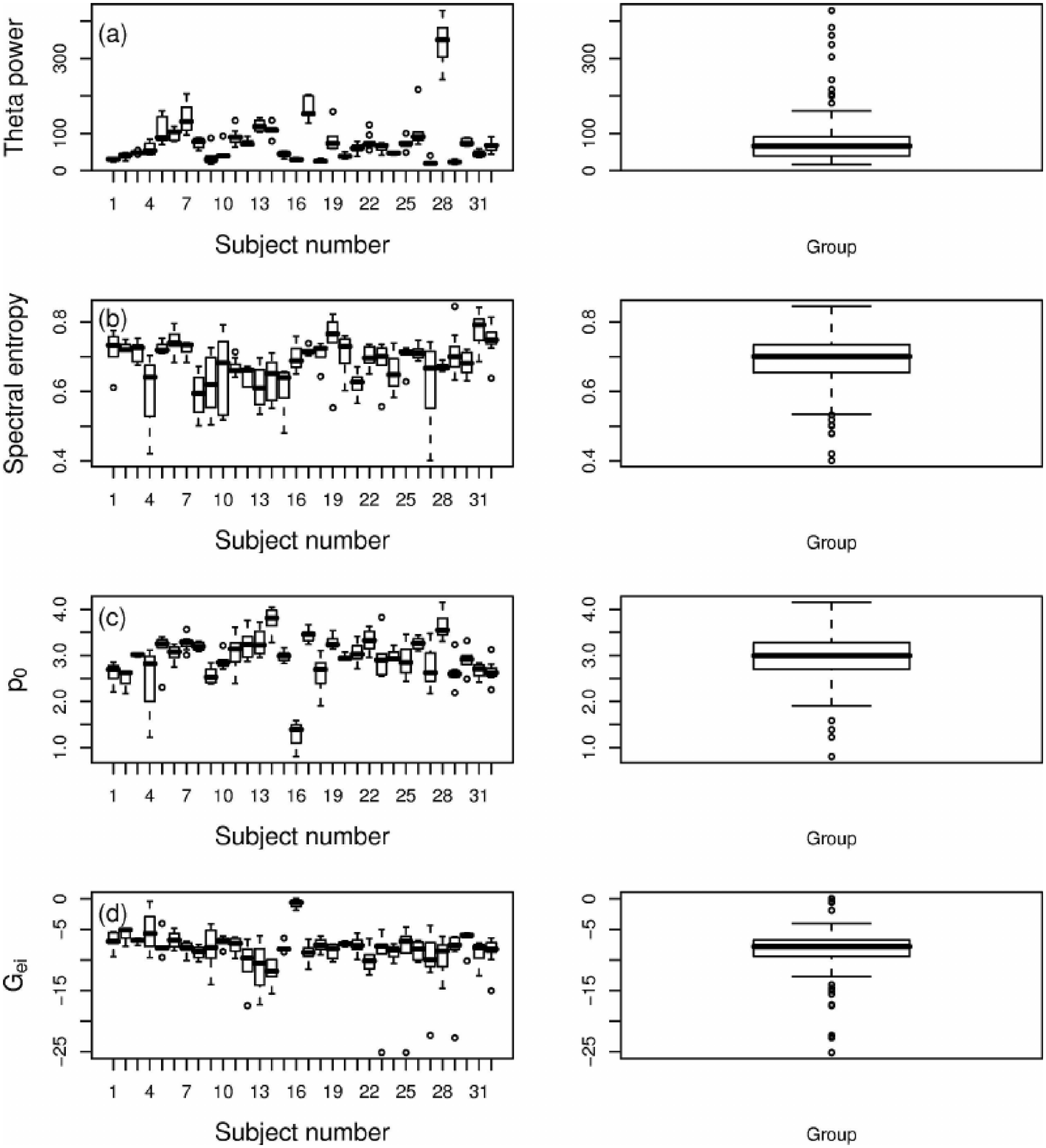}
\caption{Boxplots for each subject (left frames) and for the whole group (right frames), of (a) theta power, (b) spectral entropy, (c) overall power factor $p_0$, and (d) cortical inhibitory gain $G_{ei}$. The number of measurements per subject varies. Boxes represent the interquartile range, thick horizontal lines the median, and whiskers extend to the most extreme data points at a distance of no more than $1.5$ times the interquartile range from the box. Any observations that do no lie within the extent of the error bars are indicated by circles. The height of subjects' boxplots compared to the height of the group boxplot gives an indication of the proportion of the overall variance that is accounted for by intra-individual differences. Ratios of intra-individual to sample variance were (a) 14\%, (b) 68\%, (c) 40\%, and (d) 87\%.}
\label{fig:boxplots}
\end{figure}

The above findings may be visualized using boxplots, one for each subject, juxtaposed with a boxplot for the whole group. This has the advantage of exposing differences between subjects, and shows that some subjects are more variable than others. Boxplots for theta power, spectral entropy, $p_0$, and cortical inhibitory gain $G_{ei}$ are shown in Fig.~\ref{fig:boxplots}. Investigation of all sets of boxplots shows that a high level of variability on one measure does not imply high variability on other measures.
 
\subsection{Trial-to-trial correlations}\label{sec:correlations}

Mean Spearman rank correlations between Week 1 and Week 2, Week 2 and Week 3, etc. are given in Table~\ref{variance_ratios}. It is seen that traditional qEEG measures generally predicted future values more accurately than fitted model parameters, theta, alpha, and beta power being the most reliable ($\rho=0.83$, $\rho=0.87$, and $\rho=0.77$, respectively). These values are slightly higher than those reported by \citet{Gasser1985} for Cz spectra of children derived from 20 s of EEG taken approximately 10 months apart, but comparable to those they reported for the `best' and `worst' epoch within trials of 120 s. Their bands were defined as theta, 3.5--7.5 Hz; alpha$_1$, 7.5--9.5 Hz; alpha$_2$, 9.5--12.5 Hz; beta$_1$, 12.5--17.5 Hz; and beta$_2$, 17.5--25.0 Hz. According to their study, correlations for the 10-month interval were theta, 0.71; alpha$_1$, 0.77; alpha$_2$, 0.76; beta$_1$, 0.58; beta$_2$, 0.71; those within trials were theta, 0.87; alpha$_1$, 0.87; alpha$_2$, 0.86; beta$_1$, 0.88; beta$_2$, 0.85. We found an average rank correlation for the alpha peak frequency of $\rho=0.67$, somewhat lower than the value of 0.82 reported by Gasser et al. after a 10-month interval. The relatively low correlations for gamma ($\rho=0.50$) and delta band power ($\rho=0.45$) suggest either more true variability or a higher level of artifact in these bands than in other bands. The value of $\rho=0.45$ for the delta band is much lower than that found by Gasser et al. ($\rho=0.61$ for 10-month intervals, $\rho=0.71$ within trials), possibly due to the fact that we employed a lower frequency limit for this band, yet in both cases delta power was found to be the least reproducible of band powers. The total power had a mean trial-to-trial correlation between that of the band powers ($\rho=0.70$). The spectral entropy was the least reproducible with a mean rank correlation over the first six weeks of $\rho=0.36$. This may be compared with the test-retest reliability of $\rho=0.27$ found by Gasser et al. for a parameter counting the number of peaks in each spectrum. \citet{Kondacs1999} found an average trial-to-trial correlation after 25--62 months of 0.59 for the spectral entropy derived at 16 electrodes, but this was for frequencies up to 25 Hz. 

Of model parameters, the corticothalamic delay parameter $t_0$ had the highest trial-to-trial correlation, $\rho=0.56$. The overall power parameter $p_0$ ($\rho=0.53$) and the gain for the loop passing through the specific relay nuclei and the reticular thalamic nucleus, $G_{esre}$ ($\rho=0.50$) also showed relatively high reproducibility. The least reliable was the cortical inhibitory gain, $G_{ei}$, with a low correlation of $\rho=0.21$. 

As can be seen in Table~\ref{variance_ratios}, especially for qEEG measures, there is a strong correspondence between ratios of intra-individual variance to total variance and correlation coefficients as measures of reproducibility. This is due to the absence of consistent parameter trends in the present study. Consistent time trends across the population would leave correlation coefficients unaltered, but would increase the proportion of the population variance accounted for by intra-individual differences. 

\subsection{Intra-individual differences versus time interval}

\begin{figure}[htp]
\centering
  \includegraphics[width=350pt]{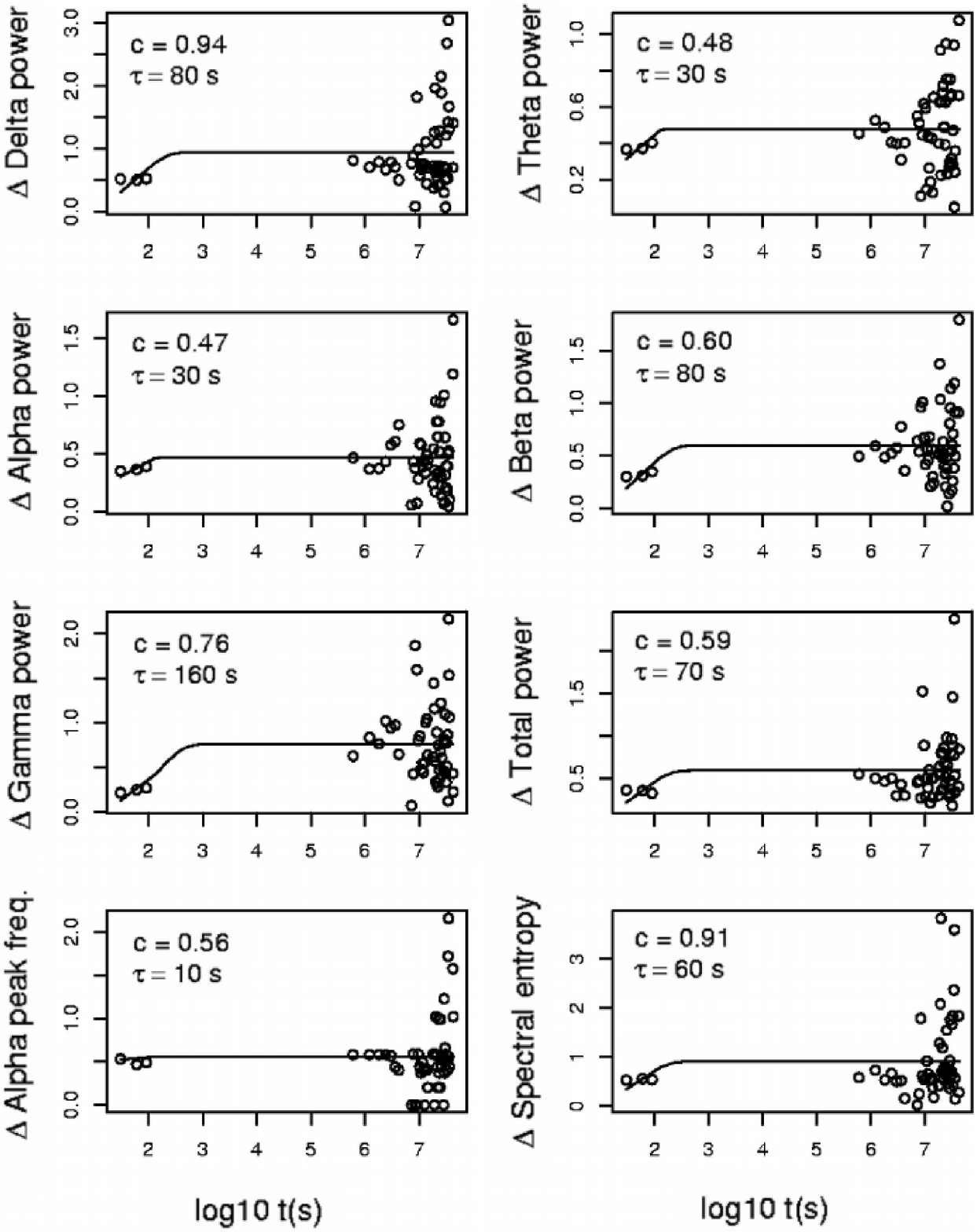}
\caption{Median absolute differences in qEEG measures versus the time interval in seconds, with fits of the form (\ref{curve}), for which the parameters $c$ and $\tau$ are given. The shorter the time scale $\tau$, the more quickly a quantity fluctuates. The given estimates are only very approximate, and data points at intervals of \mbox{$10^2$ -- $10^3$ s} are needed in order to determine time scales more precisely. The number of significant figures for time scales is one fewer than the number shown. The higher the asymptotic difference level, the less predictable a quantity becomes after a period of time. For comparison, the median absolute difference predicted for a random quantity following the standard normal distribution is around 0.95.}
\label{fig:gausstrends}
\end{figure}

\begin{figure}[htp]
\centering
  \includegraphics[width=350pt]{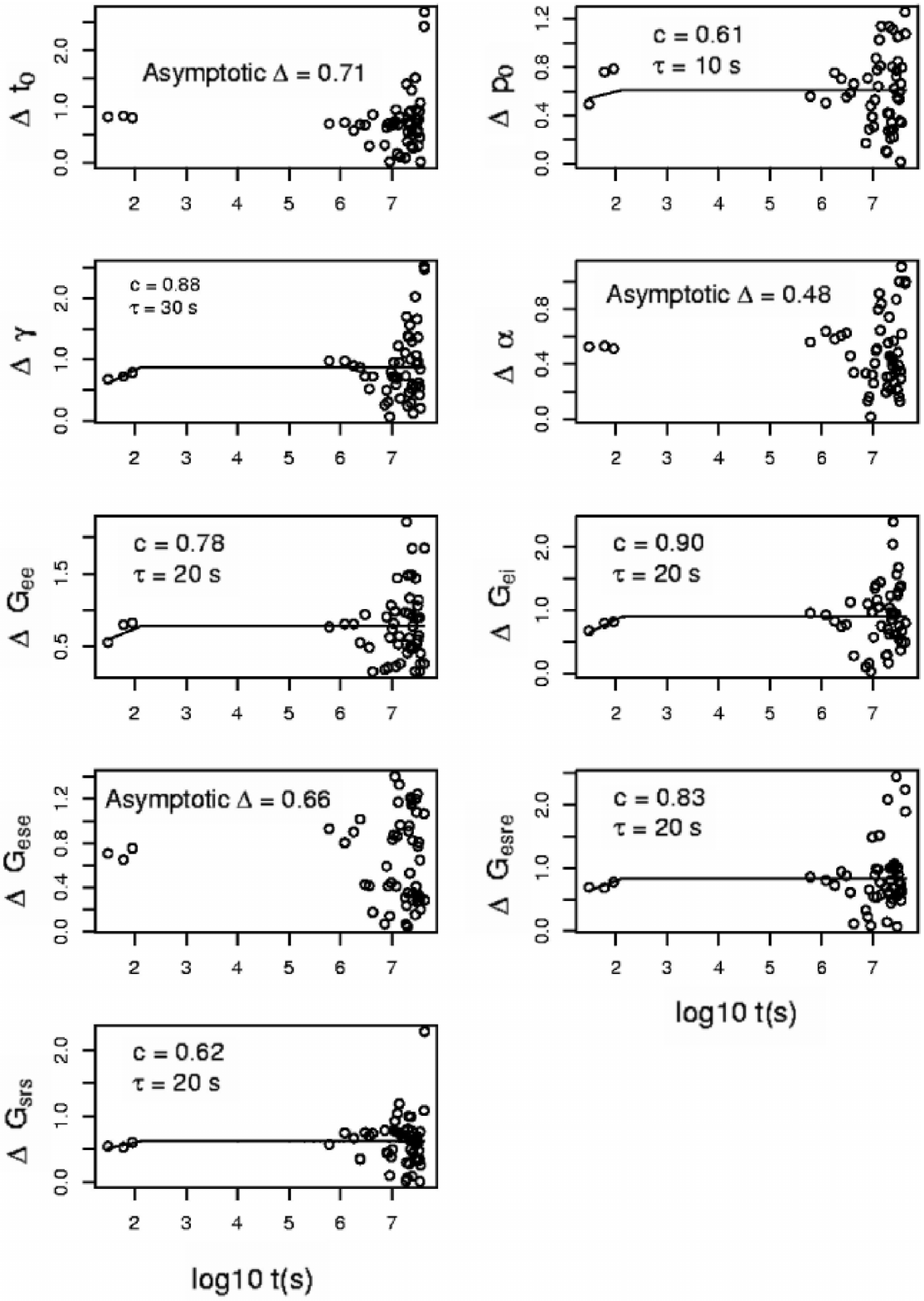}
\caption{Median absolute differences in model parameters versus the time interval in seconds, with fits of the form (\ref{curve}), for which the parameters $c$ and $\tau$ are given. For $t_0$, $\alpha$, and $G_{ese}$ only the asymptotic difference level is given, since within-trial differences were at least as large as between-trial differences. Recordings at intervals of \mbox{$10^2$--$10^3$ s} are needed to determine the time scales for fluctuations in parameters more precisely.}
\label{fig:gaussmodeltrends}
\end{figure}

 Fits of median absolute trial-to-trial differences in normalized qEEG measures and model parameters to functions of the form (\ref{curve}) are shown in Figs. \ref{fig:gausstrends} and \ref{fig:gaussmodeltrends}. 

The exponential fit decreased the residual sum of squares for all measures except $\alpha$, $G_{ese}$, and $t_0$, for which the fit did not converge, because within-trial differences were at least as large as between-trial differences. However, for these quantities we were able to estimate $c$ by taking the mean of all between-trial differences. Asymptotic differences, times after which 95\% of the variability in spectral parameters was reached, and $p$-values for the goodness of fit to curves of the form (\ref{curve}) are listed in Table~\ref{time_scales}. Since the linear correlation between median absolute difference and $c(1-e^{-t/\tau})$ was not highly significant for most quantities, and since there is a gap in observations between the scales of minutes and weeks, the results should be taken as rough estimates. Nonetheless, Table~\ref{time_scales} shows that all quantities changed on relatively short time scales of the order of minutes, suggesting that recordings of a few minutes suffice to capture the bulk of the variability in a subject's EEG.

\begin{table}[htp]
\centering
\begin{tabular}[ht]{|clccc|}
\hline
&\textbf{Quantity} & \textbf{c} & \textbf{Time scale (s)} & \textbf{\emph{p}-value} \\
\hline 
&delta power & 0.94 & 230 & 0.25 \\
&theta power & 0.48 & 80 & 0.53 \\
&alpha power & 0.47 & 90 & 0.63 \\
\textsl{qEEG} & beta power & 0.60 & 230 & 0.17 \\
&gamma power & 0.76 & 490 & 0.05 \\
&total power & 0.59 & 200 & 0.31 \\
&alpha peak freq. & 0.56 & 30 & 0.94 \\
&entropy & 0.91 & 190 & 0.46 \\
\hline 
&$t_0$ & 0.71 & -- &  --\\
&$p_0$ & 0.61 & 40 & 0.77 \\
&$\gamma$ & 0.88 & 80 & 0.65 \\
&$\alpha$ & 0.48 & -- & -- \\
\textsl{Model} & $G_{\rm{ee}}$ & 0.78 & 60 & 0.65 \\
&$G_{\rm{ei}}$ & 0.90 & 70 & 0.61 \\
&$G_{\rm{ese}}$ & 0.66 & -- & -- \\
&$G_{\rm{esre}}$ & 0.83 & 60 & 0.74 \\
&$G_{\rm{srs}}$ & 0.62 & 50 & 0.81 \\
\hline
\end{tabular}
\caption{Values of the median difference in normalized qEEG measures and model parameters after long time intervals, with approximate time scales for change defined as the time after which 95\% of the asymptotic variability has been reached. Significance values in the rightmost column refer to correlations between data points and fits of the form $\Delta(t) = c\left(1-e^{-t/\tau} \right)$, where $\Delta(t)$ is the median absolute difference between spectral parameters after a time interval $t$, $c$ is the asymptotic value of the median absolute difference, and $\tau$ is a time scale for change in the relevant parameter.}
\label{time_scales}
\end{table}


The high levels of asymptotic differences in delta power, spectral entropy, and the inhibitory cortical gain $G_{ei}$ should be compared with the level of differences expected for a completely random quantity following the standard normal distribution (0.95). Since this means that each subject's distribution becomes as wide as the population distribution, these quantities cannot be used to distinguish between individuals after intervals of more than a week. On the other hand, the decay rate of the cell-body potential ($\alpha$), the parameter that quantifies the total power ($p_0$), and the gain for the intrathalamic loop ($G_{srs}$), all had asymptotic levels of trial-to-trial differences that were comparable to those of the three most reliable band powers, theta, alpha, and beta. Comparing with Table~\ref{variance_ratios}, we find that the ratios of intra-individual to sample variance are smaller for theta, alpha, and beta band power than for $\alpha$, $p_0$, and $G_{srs}$. The difference between these two measures of reproducibility arises because the current section considers intra-subject variability as a function of the time interval between recordings, and because longer recordings were used for the analysis in Sec.~\ref{sec:ratios}.

We stress that the form of the function chosen for fitting is tentative, and that time scales cannot be determined precisely, since there is a large gap in observations for intervals of $100$ s to $10^6$ s. Rather, we provide a bound to the possible time scales for changes in EEG parameters. Therefore, in order to arrive at more precise estimates, it will be necessary to consider recordings on intermediate time scales in future work.

\section{Discussion} 

In order to estimate the significance of a difference in the electroencephalogram (EEG) between two recordings of the same individual, or between an individual's recording and the population, one needs to know the normal amounts of variability within and between individuals.

Many factors influence the reproducibility of EEG spectra, including the choice of sample; the age range, health criteria, and the homogeneity or heterogeneity of the subject group. A heterogeneous sample may have higher test-retest correlations than a homogeneous sample due to larger inter-individual differences \citep{Gasser1985}. This may have reduced the correlations in the present study, where the subject group was purposely chosen to be homogeneous.

Longer recordings reduce the amount of intra-individual variability by averaging out noise, but only up to some point. \citet{Gasser1985} found an average improvement of 0.01 for absolute power and 0.02 for relative power correlations when using 40 or 60 s instead of 20 s epochs. Also according to \citet{Salinsky1991}, correlation coefficients for 60 s epochs were slightly higher on average than coefficients based on 20 or 40 s epochs, in agreement with our finding that most, but not all, EEG variation is sampled within 20--40 s. Another reason for using short epochs is the difficulty of obtaining relatively artifact-free data over long time spans. 

Factors such as the choice of electrodes and montage are also relevant. \citet{Fein1983} found reliability of measures to be lower for temporal than for central or parietal derivations, especially in high and low-frequency bands. \citet{Salinsky1991} reported a similar result, with the T3 and T4 electrodes being less reliable on average than other electrodes. It has been found that linked ear references lead to lower test-retest correlations than a vertex reference \citep{Fein1983}. However, a vertex reference could not have been used here since the signal at the Cz electrode was considered.

Reproducibility is also intimately linked to the parameters used to quantify the spectra. Low and high-frequency bands have been consistently found to have higher intra-individual variability than intermediate bands \citep{Fein1983, Gordon2005}, in agreement with the findings in this study. Because reliability depends on the frequency, it will also depend on the band power limits employed, which differ from study to study. Similar considerations apply to determining the alpha peak frequency, for which different authors use different criteria.
Comparisons between relative and absolute band powers have been inconclusive, with some authors reporting a higher reliability for absolute power \citep{Fein1983}, others a higher reliability for relative band powers \citep{John1980}, and yet others reporting very little difference between relative and absolute band powers \citep{Salinsky1991}. \citet{Gasser1985} noted a higher reliability of relative power in the beta band but not in other bands. For the present data set, we found week-to-week correlation coefficients to be lower for relative band powers than for absolute ones, except for the delta band, which showed a slight improvement. This confirms the results of \citet{Kondacs1999}.
 
 Finally, quantification of the reproducibility itself can be done in many ways, methods in the literature including coefficients of variation, Spearman or Pearson correlations, intraclass correlations \citep{Winer1971}, and average absolute or relative differences. These may be determined between epochs or trials separated by widely different time intervals, ranging from seconds to years. It is quite difficult to estimate the uncertainty in reproducibility values due to differences in methodology and quantification, except by comparing many different studies. The findings from our study are summarized in the following paragraphs.

In a comparison between model-free and model-based spectral parameters of EEG for healthy males in the age range of 18--28 years, model parameters derived from resting eyes-closed Cz spectra for 2 minutes of EEG had a test-retest reliability that was slightly lower than that of classical qEEG measures. Fitting to a theoretical model has the complication that no model can provide a perfect description of the brain and the measuring process, and hence not all aspects of experimental spectra will be perfectly represented. This may produce intra-individual variances that are slightly higher than would be predicted on purely physiological grounds. On the other hand, fitting to a model may capture those aspects of spectra that are relatively stable within individuals by averaging out noise components. Besides, it circumvents arbitrary band power limits, and captures information on spectral shape within bands. Table~\ref{variance_ratios} shows that the first effect played a role in producing ratios of intra-individual to sample variances that were generally higher for model parameters than for qEEG measures. 

Both for qEEG measures and model parameters, transformation towards the normal distribution using a broadly applicable new method \citep{vanAlbada2007} decreased ratios of intra-individual to total sample variance when the original distribution was highly skewed, in agreement with the finding that certain skewed data have a lower test-retest reliability than non-skewed data \citep{Dunlap1994}. Spearman correlations and ratios of intra-individual to sample variance yielded very similar results due to the lack of time trends in the present study. 

Some parameters quantifying the EEG spectra or model fits to spectra were more reproducible than others. The theta and the alpha band powers were the most reproducible band powers, followed by the beta band, and finally the gamma and delta bands, which tend to be most affected by electromyographic (EMG) and electrooculographic artifact, respectively. The alpha peak frequency also showed relatively good reproducibility, which increased slightly after rejection of cases where no clear alpha peak could be discerned (from $\rho=0.67$ to $\rho=0.74$). Spectral entropy had rather poor reproducibility, in line with its dependence on vigilance (a state, rather than a trait, parameter), although artifact in the delta and gamma bands may also play a role. Of the model parameters, the conduction delay between cortex and thalamus $t_0$, and the normalization of the signal $p_0$, had the highest trial-to-trial correlations. We attribute this both to the anatomical factors determining these parameters (distance between cortex and thalamus, amount of myelination of long-range axons, skull thickness), and to the relatively straightforward relation between these parameters and spectra. The corticothalamic delay parameter $t_0$ is directly related to the alpha peak frequency, since, in our model, the alpha peak is caused by a resonance between cortex and thalamus \citep{Robinson2001}, and $p_0$ is directly related to the total power. The least reproducible model parameter was found to be $G_{ei}$, the inhibitory cortical gain. We hypothesize that this is partly due to the confounding of parameters by the fitting program in some cases, because simultaneously changing some pairs of parameters by complementary amounts leaves the spectrum virtually unchanged. Such cases would allow us to reliably fit only a combination of the parameters, rather than each parameter separately. The extent of this effect will be investigated in future studies. 

In general, reproducibility was rather low for effective comparison between individuals or groups, certainly considering that results for the Cz electrode provide a rough upper limit for the reproducibility of spectral parameters at other electrodes due to the relative absence of muscle artifact. However, increasing the sample size can improve the power of group comparisons, while repeat recordings can replace the standard deviation of a subject's score by the standard error of the mean.

Random variations increased with the time interval due to contributions from the extra sources of variability given in Table~\ref{sources_of_variability}. Based on intra-individual differences leveling off after intervals of increasing duration, as also reported by \citet{John1987} for intervals of an hour to 2.5 years, we fitted an exponential with a characteristic time scale to median absolute differences in spectral parameters. Because of a large spread in data points, as well as a large gap in observations on time scales between 1.5 minutes and 1 week, we have to be careful in interpreting our findings. For all investigated quantities, the bulk of the changes occurred with high certainty on time scales $\gtrsim$ 1 min and $\ll$ 1 week, and most probably on the scale of minutes. It is a characteristic of all physical systems (animate or inanimate) that fast changes arise from fast processes. Therefore, these relatively short time scales suggest that the variation was mainly caused by those factors in Table~\ref{sources_of_variability} that act on the scale of seconds or minutes: instrumental noise, muscle artifact, brain microstates, arousal, and attention. The latter two factors are especially implicated in fluctuations in the spectral entropy, since they act on the scale of minutes and have been linked to a broadening of the spectrum, corresponding to an increase in spectral entropy. 

Our findings thus suggest that the total level of intra-individual variation in the resting (non-drowsy) eyes-closed EEG may be estimated using recordings of only a few minutes. To establish more precisely the time scales for change in different measures, recordings at intervals intermediate between a few minutes and a week will be required in future work, where it would also be desirable to increase the number of spectra at each interval length. Time scales could then be more effectively linked to physiological factors responsible for the EEG, aided by the physiological interpretations suggested by the model. Measures that have an asymptotic level of intra-individual differences that is comparable to that of a completely random quantity cannot be used to distinguish between individuals in the long run. This is the case for delta power, spectral entropy, and the cortical inhibitory gain $G_{ei}$ in the current version of the model and fitting routine.

The spatial aspects of the model and fitting routine are being adapted to allow for multi-electrode fits, and hence to increase the amount of information that can be extracted from spectra recorded at multiple sites. This will involve modeling volume conduction and neural projection effects across the cortex, thus providing a relation between spectra at different sites. Besides leading to an improved understanding of the physiological processes responsible for the EEG, this is expected to enhance our ability to distinguish between individuals, and between normal and clinical groups, on the basis of model fits to measured cortical electrical activity. We expect that use of the methods outlined above in more extensive studies that encompass larger numbers of recordings at more diverse time scales, will greatly enhance our knowledge and understanding of the extent and sources of variability in the EEG. 
\newpage
\section*{Appendix}

The power as a function of angular frequency $\omega=2\pi f$ can be calculated after Fourier transformation of the relevant quantities, as follows \citep{Robinson2001},
%
\begin{eqnarray}\label{eq:pow}
\lefteqn{ P(\omega) = \iint |\phi_e|^2 \mathrm{exp}[-k^2/k_0^2]d^2\textbf{k} },\nonumber\\
& & {} = |\phi_n|^2 \left|\frac{G_{es}L\mathcal{P}}{1-G_{ei}L}\right|^2 \iint \frac{\mathrm{exp}[-k^2/k_0^2]}{|k^2r_e^2+q^2r_e^2|^2}d^2\textbf{k},\nonumber\\
& & {} = \frac{\pi|\phi_n|^2}{r_e^2} \left|\frac{G_{es}L\mathcal{P}}{1-G_{ei}L} \right|^2 \frac{\mathrm{Im}[\mathrm{exp}(q^{*2}/k_0^2)E_1(q^{*2}/k_0^2)]}{\mathrm{Im}~q^2r_e^2}.
\end{eqnarray}
Here, the factor $\mathrm{exp}[-k^2/k_0^2]$ is a smoothing function chosen to represent volume conduction through cerebrospinal fluid, skull, and scalp [see \citet{Robinson2001}]. The exponential integral function, $E_1$, is given by \citep{Abramowitz1970}
\begin{equation}
E_1(z) = -\gamma - \ln z - \sum_{j=1}^{\infty} \frac{(-z)^j}{jj!}
\end{equation}
for $|\arg z| < \pi$, where $\gamma = 0.5772...$ is Euler's constant.
The quantity $L$ is the dendritic filter function given by
\begin{equation}
L = (1-i\omega/\alpha)^{-1} (1-i\omega/\beta)^{-1},
\end{equation}
with $\alpha$ the decay rate and $\beta$ the rise rate for the potential at the soma. The symbol $\mathcal{P}$ represents the contribution of external input to the subcortical firing rate field,
\begin{eqnarray}
\lefteqn{\phi_s = \mathcal{P}\phi_n + \mathcal{S}\phi_e}, \nonumber\\
& & {} \mathcal{P} = \frac{L G_{sn}}{1-L^2 G_{srs}} e^{i \omega t_0/2},\nonumber\\
& & {} \mathcal{S} = \frac{L G_{se} + L^2 G_{sr}G_{re}}{1-L^2 G_{srs}}e^{i\omega t_0}.
\end{eqnarray}
Using the above expression for $\mathcal{S}$, the quantity $q^2r_e^2$ can be written as
\begin{equation}
q^2 r_e^2 = \left(1-\frac{i\omega}{\gamma_e} \right)^2 - \frac{G_{ee}L + G_{es}L \mathcal{S}}{1-G_{ei}L}.
\end{equation}
This notation is used for consistency with previous work.

The model parameter $p_0$ is contained in Eq. (\ref{eq:pow}) through
\begin{equation}\label{eq:P0eq}
P(\omega) = \frac{10^{p_0}}{G_{sn}^2}\left|\frac{L \mathcal{P}}{1-G_{ei}L} \right|^2 \frac{\mathrm{Im}[\mathrm{exp}(q^{*2}/k_0^2)E_1(q^{*2}/k_0^2)]}{\mathrm{Im}~q^2r_e^2},
\end{equation}
where the conversion between theoretical and empirical power is achieved by setting $1 \mu$V$^2$ Hz$^{-1}$m$^{-2} \equiv 1$. In other words, $p_0$ is the logarithm of the dimensionless version of 
\begin{equation}
10^{p_0} = \frac{G_{es}^2 G_{sn}^2 \pi |\phi_n(\omega)|^2}{r_e^2} \frac{\mu \mathrm{V}^2}{\mathrm{Hz}},
\end{equation}
with $r_e$ in m. Using Parseval's theorem,
\begin{equation}
\int_{-\infty}^{\infty}|f(t)|^2 dt = \int_{-\infty}^{\infty}|F(\omega)|^2 d\omega,
\end{equation}
where $F(\omega)$ is the Fourier transform of $f(t)$, we can infer that, on average, fluctuations in $|\phi_n(\omega)|^2 $ (which is dimensionless) are of the same order as those in $|\phi_n(t)|^2$ (in s$^{-2}$). This allows us to calculate approximate bounds on $p_0$. Assume $|\phi_n(\omega)|\approx 1-10$, $G_{es} \approx 0.1-2$, and $G_{sn} \approx 0.1-2$. Then 
$p_0$ lies between about $-1.3$ and $5.9$, in agreement with the values in Fig. \ref{fig:param_distr}.
 

Continuous wavenumbers $\textbf{k}$ reflect a cortex without boundaries; in the bounded case, the power becomes a sum over a discrete set of wavenumbers. If $L_x$ and $L_y$ are the linear dimensions of a rectangular cortex, the power is 
\begin{eqnarray}\label{eq:modal}
\lefteqn{ P(\omega) = |\phi_n|^2 \left|\frac{G_{es}L\mathcal{P}}{1-G_{ei}L} \right|^2 \frac{(2\pi)^2}{L_xL_y}\sum_{m,n=-M,-N}^{M,N} \frac{e^{-k_{mn}^2/k_0^2}}{|k_{mn}^2r_e^2 + q^2r_e^2|^2} }\nonumber\\
& & {} = |\phi_n|^2\left|\frac{G_{es}L\mathcal{P}}{1-G_{ei}L} \right|^2 \frac{4\pi^2}{L_xL_y}\Big[4 \sum_{m,n=1,1}^{M,N} \frac{e^{-k_{mn}^2/k_0^2}}{|k_{mn}^2r_e^2 + q^2r_e^2|^2} \nonumber\\
& & {} +2 \sum_{m=1}^M  \frac{e^{-k_{m0}^2/k_0^2}}{|k_{m0}^2r_e^2 + q^2r_e^2|^2} +2 \sum_{n=1}^N  \frac{e^{-k_{0n}^2/k_0^2}}{|k_{0n}^2r_e^2 + q^2r_e^2|^2} +\frac{1}{|q^2r_e^2|^2} \Big],
\end{eqnarray}
where the discrete wavenumbers $k_{mn}$ are given by
\begin{equation}\label{eq:modes}
k_{mn}^2 = (2\pi m/L_x)^2 +  (2\pi n/L_y)^2, 
\end{equation}
and $p_0$ is the immediate analog to that in the continuous case (\ref{eq:P0eq}). The form (\ref{eq:modal}) with (\ref{eq:modes}) is the one used in the present study, with \mbox{$L_x=L_y=$ 0.5 m}. We used $M=N=4$, so that fluctuations in the signal are taken into account over a scale of about 3 cm.


On top of the EEG spectrum, we model the EMG component by
\begin{equation}
P_{EMG}(\omega) = A_{EMG}\frac{(\omega/2 \pi f_{EMG})^2}{[1+(\omega/2 \pi f_{EMG})^2]^2},
\end{equation}
such that the EMG component has a maximum proportional to $A_{EMG}$ at about $f_{EMG}$ (taken to be 40 Hz), and tends asymptotically to $\omega^2$ at low frequencies and to $\omega^{-2}$ at high frequencies. The total spectral power density is thus \mbox{$P(\omega) + P_{EMG}(\omega)$.} The quantity $A_{EMG}$ is fitted simultaneously with the other model parameters. 

\section*{Acknowledgments}
We acknowledge the support of the Brain Resource International Database (under the auspices of The Brain Resource Company---www.brainresource.com) for use of the Eli Lilly and Modavigil datasets. We also thank the individuals who gave their time to take part in the study, and Daniel Hermens for organizing the data collection. This work was supported by an International Postgraduate Award and by the Australian Research Council.


\end{document}